\documentclass[twocolumn,showpacs,prb]{revtex4}
\newcommand{\figurewidth}{\columnwidth}
\usepackage{graphicx,amssymb}
\newcommand{\av}{_{\mathrm{av}}}

\begin{document}

\title{Large-scale Monte Carlo simulations of the three-dimensional
XY spin glass}

\author{J.~H.~Pixley}
\affiliation{Department of Physics,
University of California,
Santa Cruz, California 95064}
\author{A.~P.~Young}
\homepage{http://physics.ucsc.edu/~peter}
\email{peter@physics.ucsc.edu}
\affiliation{Department of Physics,
University of California,
Santa Cruz, California 95064}

\date{\today}

\begin{abstract}
We study the XY spin glass by large-scale Monte Carlo
simulations for sizes up to $24^3$, down to temperatures below the
transition temperature found in earlier work. The data for the larger
sizes show more marginal behavior than that for the smaller sizes
indicating that the lower critical dimension is close to, and possibly equal
to three. We find that the spins and chiralities behave in a 
similar manner. We also address the optimal ratio of ``over-relaxation''
to ``Metropolis'' sweeps in the simulation.
\end{abstract}

\pacs{75.50.Lk, 75.40.Mg, 05.50.+q}
\maketitle

\section{Introduction}

Following the convincing numerical work of Ballesteros et
al.\cite{ballesteros:00} there has been little doubt that Ising spin
glasses in three dimensions have a finite temperature transition.  In
this paper we shall study a related model for which the existence of a
finite temperature transition is more controversial: the isotropic XY
spin glass, which is composed of classical spins with two components.
Early work on this model in three
dimensions\cite{morris:86,jain:86} indicated a zero temperature
transition, or possibly a transition at a very low but non-zero
temperature.  However, following the pioneering work of
Villain\cite{villain:77b}, which emphasized the role of ``chiralities''
(Ising-like variables which describe the handedness of the non-collinear
spin structures), Kawamura and Tanemura\cite{kawamura:87} proposed that
the spin glass transition only occurs at $T_{SG} = 0$ and that a
\textit{chiral} glass transition occurs at a finite temperature
$T_{CG}$.  This scenario requires that spins and chiralities
\textit{decouple} at
long length scales.  Kawamura and collaborators have given numerical
results in favor of
this scenario\cite{kawamura:01}.

However, the absence of a spin glass transition in the XY spin glass has
been challenged by Maucourt and Grempel\cite{maucourt:98} and
subsequently Akino and Kosterlitz\cite{akino:02} who found evidence for
a possible finite $T_{SG}$ from zero temperature domain wall
calculations.  Furthermore, by studying the dynamics of the XY spin
glass in the phase representation, Granato\cite{granato:00} found that
the ``current-voltage'' characteristics exhibited scaling behavior which
he interpreted as a transition in the spins as well as the chiralities.

In earlier work\cite{leeLW:03}, referred to as LY, Lee and one of the
present
authors studied spin and chiral correlations on an equal footing, using
the method of analysis that was the most successful for the Ising spin
glass\cite{ballesteros:00,katzgraber:06,hasenbusch:08},
namely finite-size scaling of the
correlation length. Considering a modest range of sizes, $N = L^3$ with
$L \le 12$, LY
found that the behavior of spins and chiralities was quite similar and
they both had a finite temperature transition, apparently at the same
temperature.

LY studied both XY and Heisenberg models, finding similar conclusions
for both. However, for the Heisenberg case, subsequent studies on much
larger sizes\cite{campos:06,lee:07}, up to $L=32$, have painted a
more complex picture.  The data at the lowest temperatures and
largest sizes seems rather ``marginal'', i.e.~the system is close to the
lower critical dimension where the finite-temperature phase transition
is removed by fluctuations.  The data for spins and chiralities are
still quite similar, though not identical, and do not seem to give
compelling evidence for spin-chirality decoupling as proposed by
Kawamura. In addition, Hukushima and Kawamura\cite{hukushima:05} have
also studied somewhat larger sizes than LY ($L \le 20$),
but they argued that their
data \textit{is} consistent with spin-chirality decoupling.

It is of interest to know whether the ``crossover'' to more marginal
behavior found for larger sizes is special to the three-component case,
or whether the same situation occurs quite generally with vector spin
glasses. In this paper, we therefore study the XY (2-component) spin
glass for larger sizes (up to $24^3$) than in LY (which went only up to
$12^3$). We find a situation that is quite similar to the Heisenberg
case, namely marginal behavior for low-$T$ and large sizes. The behavior
of the spin glass and chiral glass correlation length
is \textit{very} similar, more similar than
was the case for the Heisenberg spin glass, and 
does not appear to provide
evidence for
spin-chirality decoupling, at least up to the sizes studied.

Simulations on very large sizes for vector spin glasses have been
possible because including
``overrelaxation'' moves, in addition to the more familiar Metropolis or
heatbath moves, speeds up equilibration~\cite{alonso:96}.
A second motivation of the
present work is to investigate quantitatively the \textit{optimal} ratio of
overrelaxation to Metropolis sweeps for the XY spin glass. 

The layout of this paper is as follows. Section~\ref{sec:manda}
describes the model, the parameters of the simulations, and the
finite-size scaling approach. The results for the correlation length
are presented in
Sec.~\ref{sec:res}. In Sec.~\ref{sec:opt_equil} we estimate
the optimal ratio between the number
of overrelaxation and Metropolis sweeps, and
Sec.~\ref{sec:concl} summarizes our conclusions.

\section{Model and analysis}
\label{sec:manda}

We use the standard Edwards-Anderson XY spin glass model
\begin{equation}
{\cal H} = -\sum_{\langle i, j \rangle} J_{ij} {\bf S}_i \cdot
{\bf S}_j,
\end{equation}
where the ${\bf S}_i$ are 2-component classical
vectors of unit length at the sites of a simple cubic lattice, and the
$J_{ij}$ are nearest neighbor interactions with a Gaussian distribution
with zero mean and standard deviation unity.  Periodic boundary
conditions are applied on lattices with $N=L^3$ spins. 

The spin glass order parameter, $q^{\mu\nu}({\bf k})$, at
wave vector ${\bf k}$, is defined to be
\begin{equation}
q^{\mu\nu}({\bf k}) = {1 \over N} \sum_i S_i^{\mu(1)} S_i^{\nu(2)}
e^{i {\bf k} \cdot {\bf R}_i},
\end{equation}
where $\mu$ and $\nu$ are spin components, and ``$(1)$'' and ``$(2)$''
denote two
identical copies of the system with the same interactions. From this we
determine the wave vector dependent
spin glass susceptibility $\chi_{SG}({\bf k})$ by
\begin{equation}
\chi_{SG}({\bf k}) = N \sum_{\mu,\nu} [\langle \left|q^{\mu\nu}({\bf
k})\right|^2 \rangle ]\av ,
\end{equation}
where $\langle \cdots \rangle$ denotes a thermal average and
$[\cdots ]\av$ denotes an average over disorder. The spin glass correlation
length
is then determined\cite{palassini:99b,ballesteros:00} from 
\begin{equation}
\xi_L = {1 \over 2 \sin (k_\mathrm{min}/2)}
\left({\chi_{SG}(0) \over \chi_{SG}({\bf k}_\mathrm{min})} - 1\right)^{1/2},
\end{equation}
where ${\bf k}_\mathrm{min} = (2\pi/L)(1, 0, 0)$.

For the XY spin glass, chirality of a square is~\cite{kawamura:01}
\begin{equation}
\label{kappa_mu}
\kappa_i^\mu = {1 \over 2\sqrt{2}} \sum_{\langle l,m
\rangle}^{\hspace{7mm}\prime}
\mathrm{sgn}(J_{l m}) \sin (\theta_l - \theta_m),
\end{equation}
where $\theta_l$ is the angle characterizing the direction of spin
${\bf S}_l$, and the prime on the sum
indicates that it is over the four bonds around the elementary
plaquette
perpendicular to the $\mu$ axis whose ``bottom left'' corner is
site $i$.
The chiral glass susceptibility is then given by
\begin{equation}
\label{chisg}
\chi_{CG}^\mu({\bf k}) =  N [\langle \left| q_{c}^\mu({\bf k})\right|^2
\rangle ]\av ,
\end{equation}
where the chiral overlap $q_{c}^\mu({\bf k})$ is given by
\begin{equation}
\label{qc}
q_{c}^\mu({\bf k}) = {1 \over N} \sum_i  \kappa_i^{\mu(1)} \kappa_i^{\mu(2)}
e^{i {\bf k} \cdot {\bf R}_i}.
\end{equation}
We define
the chiral correlation lengths $\xi^\mu_{c,L}$ by
\begin{equation}
\label{xi_c}
\xi^\mu_{c,L} = {1 \over 2 \sin (k_\mathrm{min}/2)}
\left({\chi_{CG}(0) \over \chi_{CG}^\mu({\bf k}_\mathrm{min})} - 1\right)^{1/2},
\end{equation}
in which $\chi_{CG}({\bf k}=0)$ is independent of $\mu$.
Note that $\xi^\mu_{c,L}$ will, in general, be different for $\hat{\mu}$ along
${\bf k_\mathrm{min}}$ (the $\hat{x}$ direction)
and perpendicular to ${\bf k}$, though this difference is very
small for large sizes. The results presented will be an average over
the three (two transverse and one longitudinal) correlation lengths.

To equilibrate the system efficiently we perform three types of Monte
Carlo move.

%\begin{enumerate}
%\item
Firstly we use ``over-relaxation'' sweeps\cite{alonso:96}
%(also known as ``over-relaxation'' sweeps).
in which we sweep sequentially through the lattice, and, at
each site, compute the local field on the spin, $\mathbf{H}_i = \sum_j J_{ij}
\mathbf{S}_j$. The new value for the spin on site $i$ is taken to be
its old value reflected about $\mathbf{H}$, i.e.
\begin{equation}
\mathbf{S}'_i = -\mathbf{S}_i + 2\, {\mathbf{S}_i \cdot \mathbf{H}_i \over
H_i^2}\, \mathbf{H}_i \, .
\label{reflect}
\end{equation}
%see Fig,~\ref{micro}.
Over-relaxation sweeps 
preserve energy and so are also known as microcanonical sweeps.
%They are very fast because the
%operations are simple and no random numbers are needed. Furthermore, as
%discussed in Sec.~\ref{sec:opt_equil} including over-relaxation sweeps
%actually \textit{reduces} the total number of sweeps needed to
%equilibrate.

Secondly, we
include Metropolis sweeps since, unlike the over-relaxation sweeps,
these \textit{do} change the energy, and so are needed to bring the system
to equilibrium.
For the data presented in Secs.~\ref{sec:manda} and \ref{sec:res}
we do one Metropolis sweep
after every 10 over-relaxation sweeps.  As for the
over-relaxation case, we sweep sequentially through the lattice. To
update a given spin, we choose a trial new direction randomly within a window
$\pm \Delta \theta/2$ of the current direction, and accept this new
direction with the usual Metropolis probability, $\mbox{min}(1,
\exp(-\beta \Delta E))$,
where $\beta = 1/T$ and $\Delta E$ is the energy difference between
the trial state and the current state. We choose the window size $\Delta
\theta$ to vary with temperature in such a way that the acceptance ratio for
Metropolis moves is in the range of 30 to 50\%.

A Metropolis sweep requires more CPU time than a over-relaxation sweep,
so we do mainly over-relaxation sweeps, including \textit{some} Metropolis
sweeps only to change the energy from time to time
to ensure that the algorithm is ergodic.
In fact, as discussed in Sec.~\ref{sec:opt_equil},
including 
a fraction of over-relaxation sweeps 
not only \textit{reduces the CPU time} (for a given
\textit{total} number of sweeps) but also \textit{reduces the 
number of sweeps} needed to equilibrate. 

Finally we do ``parallel tempering'' sweeps
\cite{hukushima:96,marinari:98b}, which are necessary to prevent the
system being trapped in a valley in configuration space at low
temperatures.  One takes $N_T$ copies of the system with the same bonds
but at a range of different temperatures.  The minimum temperature,
$T_{\rm min} \equiv T_1$, is the low temperature where one wants to
investigate the system (below $T_{SG}$ in our case), and the maximum,
$T_{\rm max} \equiv T_{N_T}$, is high enough that the the system
equilibrates very fast (well above $T_{SG}$ in our case). A parallel
tempering sweep consists of swapping the temperatures of the spin
configurations at a pair of neighboring temperatures, $T_i$ and
$T_{i+1}$, for $i = 1, 2, \cdots , T_{N_T - 1}$ with a probability that
satisfies the detailed balance condition. Further details on the
application to vector spin glasses can be found in
Ref.~\onlinecite{lee:07}. For the simulations in Secs.\ref{sec:manda}
and
\ref{sec:res} we do one parallel tempering sweep after each Metropolis
sweep.

Table \ref{simparams} gives the parameters of the simulations used to
collect the data in Secs.~\ref{sec:manda} and \ref{sec:res}.

\begin{table}
\caption{
Parameters of the simulations described in Secs.~\ref{sec:manda} and
\ref{sec:res}.  $N_{\rm samp}$ is the number of samples,
$N_{\rm equil}^{OR}$ is the number of over-relaxation Monte Carlo sweeps for
equilibration for each of the $2 N_T$ replicas for a single sample, and
$N_{\rm meas}^{OR}$ is the number of over-relaxation sweeps for measurement.
The number of Metropolis sweeps and the number of parallel tempering
sweeps are both equal to 10\% of the number of
over-relaxation sweeps.  $T_{\rm min}$ and $T_{\rm max}$ are the
lowest and highest temperatures simulated,
and $N_T$ is the number of temperatures
used in the parallel tempering.
\label{simparams}
}
\begin{tabular*}{\columnwidth}{@{\extracolsep{\fill}} r r r r r r r } 
\hline
\hline
$L$  & $N_{\rm samp} $ & $N_{\rm equil}^{OR}$ &
$N_{\rm meas}^{OR}$ & $T_{\rm min}$ & $T_{\rm max}$ & $N_{T}$  \\ 
\hline
 4 &  5000 & 1280    & 1280    & 0.200 & 1.40 & 11 \\
 6 &  5001 & 10240   & 10240   & 0.200 & 1.40 & 19 \\
 8 &  1000 & 40960   & 40960   & 0.200 & 1.40 & 27 \\
12 &  1000 & 81920   & 81920   & 0.250 & 0.60 & 24 \\
16 &  1006 & 409600  & 409600  & 0.265 & 0.60 & 32 \\ 
24 &  461  & 2457600 & 2457600 & 0.265 & 0.45 & 35 \\ 
\hline
\hline
\end{tabular*}
\end{table}

\begin{figure}[!tbp]
\begin{center}
\includegraphics[width=\figurewidth]{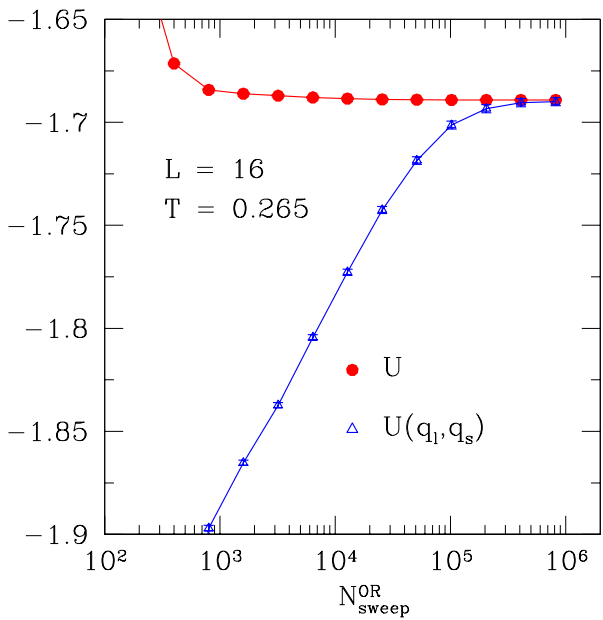}
\end{center}
\caption{(Color online)
Equilibration plot, testing Eq.~(\ref{equiltest}),
for $L=16$ at $T=0.265$. It is seen that
the data for $U$ and $U(q_l, q_s)$, given by Eq.~(\ref{Uqlqs}),
come together when the
total number of over-relaxation 
sweeps, $N_{\rm sweep}^{OR} = N_{\rm equil}^{OR} + N_{\rm meas}^{OR}$,
see Table \ref{simparams},
is equal to about
$2 \times 10^5$. These two quantities then
stay at their common value
indicating that equilibration has been achieved.
%The lines are guides to the eye.
It is seen that the
energy comes close to its equilibrium value very quickly,
whereas $U(q_l, q_s)$, which depends on the link overlap $q_l$ between
two replicas,
takes much longer.
\label{equil}
}
\end{figure}

\begin{figure}[!tbp]
\begin{center}
\includegraphics[width=\figurewidth]{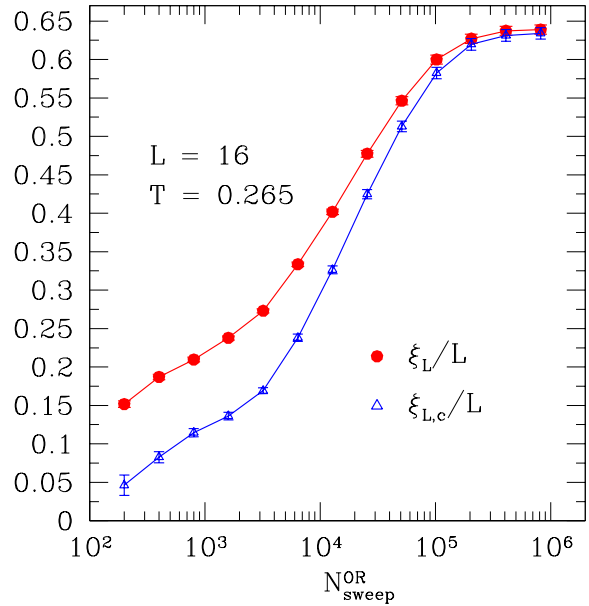}
\end{center}
\caption{(Color online)
A plot of the spin glass and chiral glass correlation lengths, $\xi_L$
and $\xi_{L, c}$, divided by
$L$, as a function of the total number of sweeps for
$L=16$ at $T=0.265$. It is seen the data flattens
off at around $2 \times 10^5$ sweeps, the value where the two sets of
data in Fig.~\ref{equil} start to agree. This indicates that when the
data in Fig.~\ref{equil} agree within high precision, i.e.~when
Eq.~(\ref{equiltest}) is satisfied, the correlation lengths have reached
their equilibrium value.
%The lines are guides to the eye.
\label{equil_xi} 
}
\end{figure}

To test for
equilibration\cite{katzgraber:01} we require that data satisfy the 
relation\cite{lee:07}
\begin{equation}
U = U(q_l, q_s)
\label{equiltest}
\end{equation}
where
\begin{equation}
U(q_l, q_s) = 
{z\over 2 T}\, 
(q_l - q_s ) \, ,
\label{Uqlqs}
\end{equation}
which is valid for a Gaussian bond distribution. Here
$U = - [\sum_{\langle i, j \rangle} J_{ij} \langle {\bf S}_i \cdot
{\bf S}_j \rangle ]\av $ is
the average energy per spin, $q_l = (1/N_b)\sum_{\langle i, j \rangle}
[ \langle
{\bf S}_i \cdot {\bf S}_j \rangle^2]\av$ is the ``link overlap'', $q_s =
(1/N_b)\sum_{\langle i, j \rangle}[\langle ({\bf S}_i \cdot {\bf S}_j)^2
\rangle]\av$, $N_b = (z/2)N$ is the number of nearest neighbor bonds, and
$z\ (=6\ \mbox{here})$ is the lattice coordination number. Equation
(\ref{equiltest}) is easily derived by integrating by parts the
expression for the average energy with respect to $J_{ij}$, noting that the
average $[\cdots]\av$ is over a Gaussian function of the
$J_{ij}$'s.

The spins are initialized in random directions so the energy, the LHS of
Eq.~(\ref{equiltest}), is initially close to zero and decreases,
presumably monotonically, to its equilibrium value as the length of the
simulation increases. Hence the LHS of Eq.~(\ref{equiltest}) will be too
\textit{large} if the simulation is too short to equilibrate the system.
On the other hand, the RHS of Eq.~(\ref{equiltest}), will be too
\textit{small} if the simulation is too short because $q_l$ starts off
close to zero and then increases with MC time as the two replicas start
to find the same local minima. The quantity $q_s$ will be less dependent
on Monte Carlo time than $q_l$ since it is a local variable for a single
replica.  (For the Ising case it is just a constant.) Hence if the
simulation is too short the RHS of Eq.~(\ref{equiltest}) will be too
low. In other words, the two sides of Eq.~(\ref{equiltest}) 
are expected to approach the
common equilibrium value from \textit{opposite directions} as the length
of the simulation increases.
Only if Eq.~(\ref{equiltest}) is satisfied within small error
bars do we accept the results of a simulation.

Figure \ref{equil} shows a test to verify that Eq.~(\ref{equiltest}) is
satisfied at long times. For the parameters used, $L=16, T=0.265$, this
occurs when the total number of (over-relaxation)
sweeps ($N_{\rm sweep}^{OR}=N_{\rm equil}^{OR}+N_{\rm meas}^{OR}$)
is about $2 \times 10^5$. Figure
\ref{equil_xi} shows that the spin
and chiral correlation lengths appear to become independent of $N_{\rm sweep}$,
and hence are presumably equilibrated, when $N_{\rm sweep}$ is larger than this
\textit{same} value. Hence, it appears that when Eq.~(\ref{equiltest}) is
satisfied to high precision, the data for the correlation lengths is
equilibrated.

With the number of sweeps shown in Table \ref{simparams},
Eq.~(\ref{equiltest}) was satisfied for all sizes and temperatures. The
error bars are made sufficiently small by averaging over a large number
of samples.

Since $\xi_L/L$ is dimensionless it has the finite size scaling
form\cite{ballesteros:00,palassini:99b,leeLW:03}
\begin{equation}
{\xi_L \over L} = \widetilde{X}\left(L^{1/\nu}(T - T_{SG})\right) ,
\label{eq:fss}
\end{equation}
where $\nu$ is the correlation length exponent.  Note that there is no
power of $L$ multiplying the scaling function $\widetilde{X}$. By contrast,
for the spin glass susceptibility, $\chi_{SG} \equiv \chi_{SG}(\textbf{k}=0)$,
which has dimensions,
the finite-size scaling form is 
\begin{equation}
\chi_{SG} = L^{2-\eta_{SG}}\, \widetilde{K}\left(L^{1/\nu}(T - T_{SG})\right) ,
\label{eq:fss-chisg}
\end{equation}
where $\eta_{SG}$ is a critical exponent.
There is an 
expression analogous to Eq.~(\ref{eq:fss})
for the chiral correlation length, and to Eq.~(\ref{eq:fss-chisg})
for the chiral glass
susceptibility $\chi_{CG} \equiv \chi_{CG}(\textbf{k}=0)$. For the later case,
there is no reason to expect that the exponents $\eta_{SG}$ and $\eta_{CG}$
are equal.

From Eq.~(\ref{eq:fss})
it follows that the data for $\xi_L/L$ for
different sizes come together at
$T=T_{SG}$. In addition, they are also expected to splay out again on
the low-$T$ side~\cite{ballesteros:00}
if there is spin glass order below $T_{SG}$. In a
marginal situation with a line of critical points, as in the
Kosterlitz-Thouless-Berezinskii theory of the transition in the
two-dimensional XY ferromagnet,
the data for different sizes would come together at $T_{SG}$
and then stick together at lower $T$, see for example Fig.~3 of
Ref.~\onlinecite{ballesteros:00}.

\begin{figure}[!tbp]
\begin{center}
\includegraphics[width=\figurewidth]{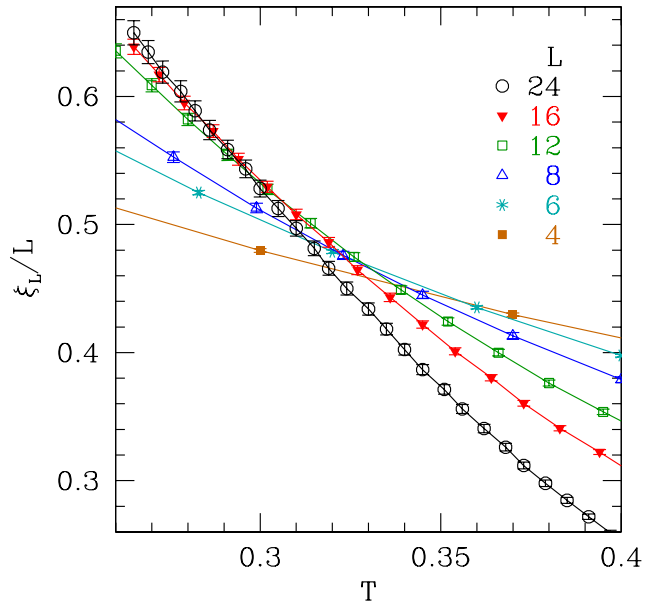}
\end{center}
\caption{(Color online)
Data for $\xi_L/L$,
the spin glass correlation length divided by system size,
as a function of $T$ for different system sizes. 
\label{xi_L} 
}
\end{figure}

\begin{figure}[!tbp]
\begin{center}
\includegraphics[width=\figurewidth]{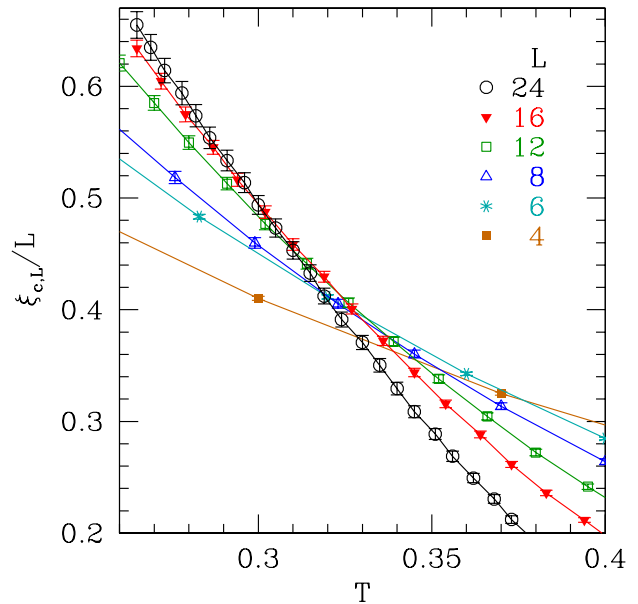}
\end{center}
\caption{(Color online)
Data for the chiral correlation length (averaged over longitudinal and
transverse directions)
divided by system size,
as a function of $T$ for different system sizes.
\label{xichiral_L} 
}
\end{figure}

%\begin{figure}[!tbp]
%\begin{center}
%%\includegraphics[width=\figurewidth]{xi_chiralx_L_enlarge.eps}
%\end{center}
%\caption{(Color online)
%Enlarged view of a region of Fig.~\ref{xichiral_L}.
%\label{xi_chiral_L_enlarge} 
%}
%\end{figure}

\section{Results}
\label{sec:res}

We studied sizes from $L=4$ to $L=24$, as shown in Table~\ref{simparams}.
The CPU time involved to get this data is about
8 Mac G5 CPU years.

The data for the spin glass correlation length (divided by $L$) is shown in
Fig.~\ref{xi_L}, and the corresponding data
for the chiral glass correlation length is shown in
Fig.~\ref{xichiral_L}. In both cases the data for smaller sizes
intersect and splay out at lower temperature. However, for the larger
sizes the splaying out is small, indicating close to ``marginal''
behavior, i.e. the ``lower critical dimension'' is close to 3.

\begin{figure}[!tbp]
\begin{center}
\includegraphics[width=\figurewidth]{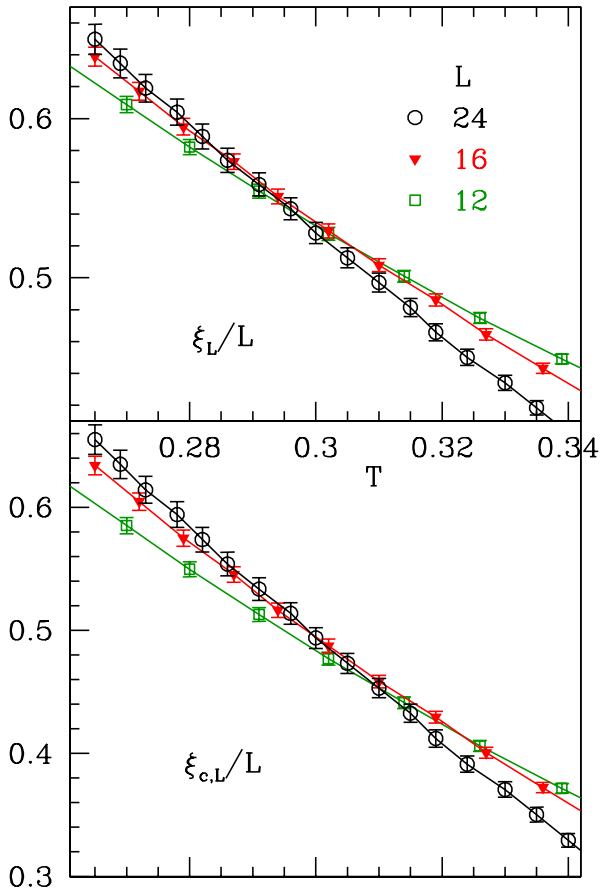}
\end{center}
\caption{(Color online)
The same data as in Figs.~\ref{xi_L} and \ref{xichiral_L},
but including only the largest sizes and
in a somewhat expanded scale.
\label{xi_L_both} 
}
\end{figure}

The data for the spins and chiralities in Figs.~\ref{xi_L} and
Fig.~\ref{xichiral_L}, are very similar, so we do not see evidence
for spin-chirality decoupling. To make clearer the similarity between
the two sets of data we plot them both in Fig.~\ref{xi_L_both},
including just the three largest sizes. The temperature where the data
merges decreases slightly with increasing size. We have estimated the
temperatures where the data intersect/merge for different pairs of sizes and
present the results in Table~\ref{crossings}.
The temperatures are seen to decrease with increasing
size. If one neglects the smallest pair of sizes ($L=4/6$) the shift is
somewhat bigger for the spins than for the chiralities but,
from the data, it is not possible to reliably estimate whether or not the
intersection temperature will tend to zero for $L \to\infty$ for either set of data.

\begin{table}
\caption{
Estimated crossing temperatures
for the spin and chiral glass correlation lengths. The
results are given to the nearest $0.005$, but the uncertainties are greater
than this because of the error bars in the data itself.
\label{crossings}
}
%\begin{tabular*}{\columnwidth}{@{\extracolsep{\fill}} r r r r r r r } 
\begin{tabular*}{\columnwidth}{@{\extracolsep{\fill}} l l l } 
\hline
\hline
sizes & $T_{\rm crossing}$ (spins) &  $T_{\rm crossing}$ (chiralities) \\ 
\hline
 4/6  &  0.355 & 0.375 \\
 6/8  &  0.33  & 0.32  \\
 8/12 &  0.33  & 0.335 \\
12/16 &  0.31  & 0.32  \\
16/24 &  0.285 & 0.30  \\ 
\hline
\hline
\end{tabular*}
\end{table}

\begin{figure}[!tbp]
\begin{center}
\includegraphics[width=\figurewidth]{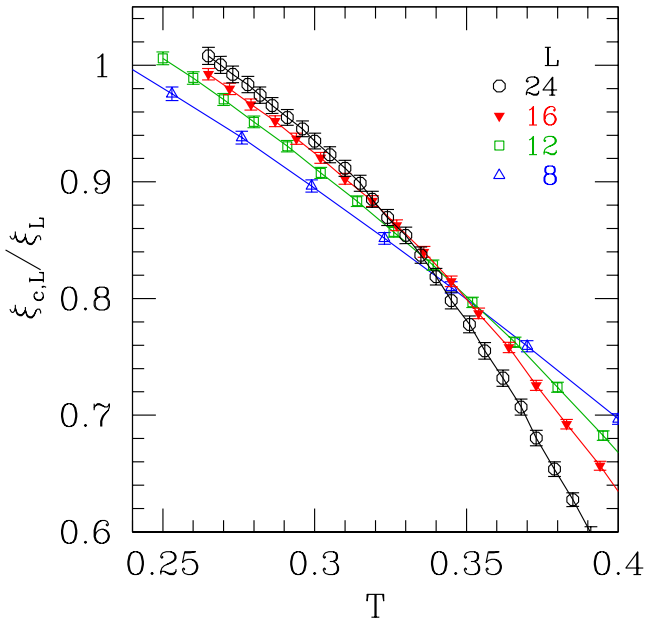}
\end{center}
\caption{(Color online)
Data for the \textit{ratio}
of the chiral glass to the spin glass correlation
lengths for sizes from 8 to 24.
\label{fig:chi_ratio} 
}
\end{figure}

In Fig.~\ref{fig:chi_ratio} we present data for the \textit{ratio}
of the chiral-glass to spin-glass correlation lengths. For the largest
sizes the data intersects for $T$ about 0.33 and then (slightly) splays
out in the low-$T$ side.
If there is a single transition involving both
spins and chiralities, then the data would become independent of size at
the transition (since both $\xi_L$ and $\xi_{c,L}$ are proportional to
$L$
there, see Eq.~(\ref{eq:fss})).
If the stiffness exponents for spins and chiralities are
equal (we are not aware of any argument for this even if there is a
single transition) then the data would become independent of $L$ for
large $L$ at low-$T$. If the stiffness exponent for chiralities is
larger than that for the spins, then the ratio would diverge in this
limit.  From the data it is not possible to say for sure if the data
diverges or not at low-$T$, but the size dependence at the larger sizes
is very weak.

In the spin-chirality decoupling scenario, the ratio would diverge even
at the transition, and there would not be a common intersection.
We feel that the data of Fig.~\ref{fig:chi_ratio} reinforces our view
that if spin-chirality decoupling occurs one would need even larger
sizes than $L=24$ to see it.

\begin{figure}[!tbp]
\begin{center}
\includegraphics[width=\figurewidth]{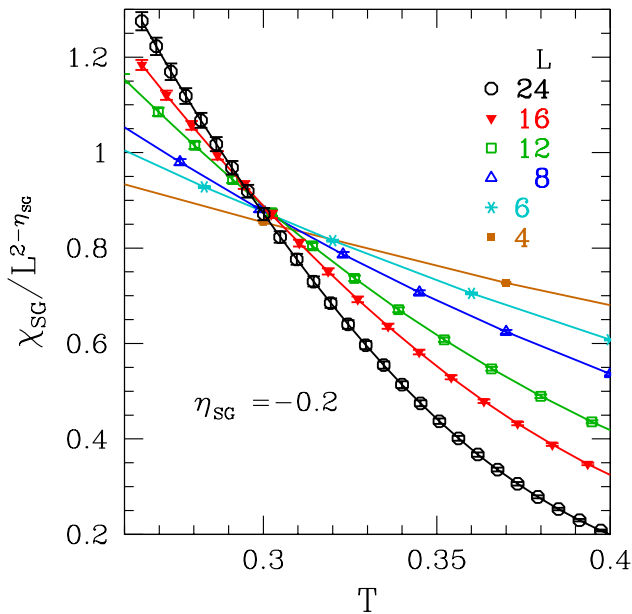}
\end{center}
\caption{(Color online)
Data for the spin glass susceptibility $\chi_{\rm SG} \equiv \chi_{\rm
SG}(\mathbf{k}=0)$ divided by $L^{2-\eta_{SG}}$ where we took
$\eta_{SG} = -0.2$ in
order to get the data to intersect (see Eq.~(\ref{eq:fss-chisg}))
for $T$ around 0.30 since this is roughly where the
data for $\xi_L/L$ and $\xi_{c,L}$ intersect/merge
for the largest sizes, see
Figs.~\ref{xi_L} and \ref{xichiral_L}.
\label{fig:chisg} 
}
\end{figure}

\begin{figure}[!tbp]
\begin{center}
\includegraphics[width=\figurewidth]{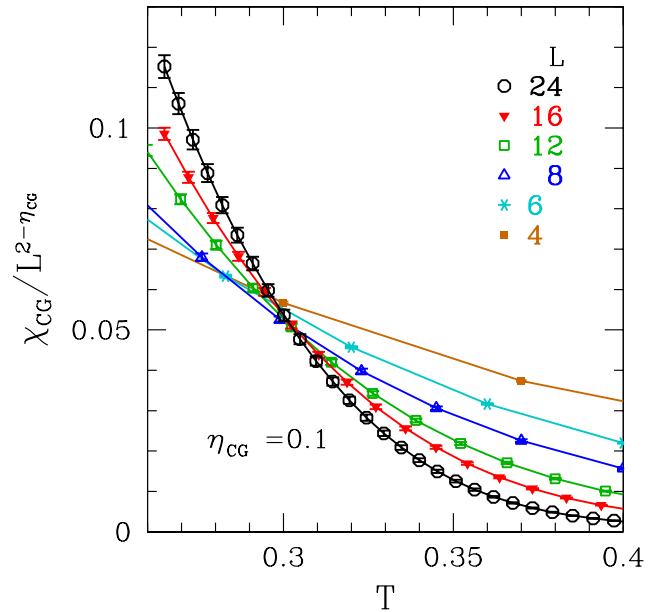}
\end{center}
\caption{(Color online)
Similar to Fig.~\ref{fig:chisg} but for the
chiral glass susceptibility $\chi_{\rm CG} \equiv \chi_{\rm
CG}(\mathbf{k}=0)$.
Here we took $\eta_{CG} = 0.1$.
\label{fig:chicg} 
}
\end{figure}

We also present data for the spin-glass and chiral-glass susceptibilities in
Figs.~\ref{fig:chisg} and \ref{fig:chicg} respectively. Dividing by $L^{2 -
\eta}$, where $\eta$ is a critical exponent, the data should intersect at the
critical temperature, see Eq.~(\ref{eq:fss-chisg}),
where $\eta_{SG}$ is not necessarily
equal to $\eta_{CG}$. In order to get intersections for $T \simeq 0.30$, where
the correlation data merge/intersect for the largest sizes, we took $\eta_{SG}
= -0.2$ and $\eta_{CG} = 0.1$ in the plots.

Given the large corrections to scaling clearly visible in the data for
the correlation lengths,
it does not
appear possible to get \textit{reliable} estimate of the critical exponents,
$\eta_{SG}$ and $\eta_{CG}$, or of the correlation length exponent $\nu$.

\section{Optimizing the fraction of overrelaxation sweeps}
\label{sec:opt_equil}

\begin{figure}[!tbp]
\begin{center}
\includegraphics[width=\figurewidth]{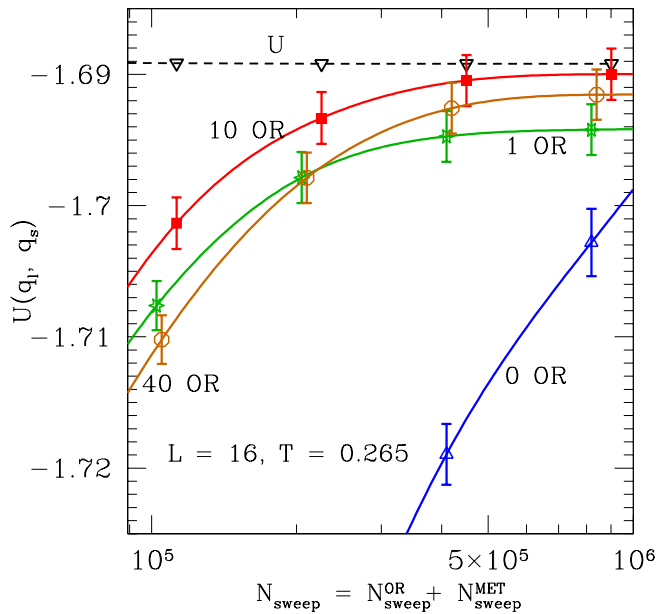}
\end{center}
\caption{(Color online)
Results for $L=16, T = 0.265$.
The data connected by solid lines is $U(q_l, q_s)$ in Eq.~(\ref{Uqlqs}) for
different number of over-relaxation (OR) sweeps per Metropolis (MET) sweep as
indicated.
The horizontal axis is the total number of OR plus MET sweeps.
The data connected by the dashed line is the energy $U$,
which should equal $U(q_l, q_s)$ in equilibrium according to
Eq.~(\ref{equiltest}).
Since the energy equilibrates relatively fast, its value
does not depend significantly on the ratio of OR to MET sweeps for the
range of sweeps presented. The number of parallel tempering sweeps is
the same for all sets of data except for ``40 OR'' where it is $1/4$ as
many.
\label{opt_equil} 
}
\end{figure}

As already noted, adding overrelaxation steps has been
observed~\cite{alonso:96,campos:06,lee:07} to speed up equilibration.
Here we look systematically at how the \textit{ratio} of the number of
over-relaxation (OR) sweeps to Metropolis (MET) sweeps alters the
\textit{total} number of sweeps needed to equilibrate. In
Fig.~\ref{opt_equil}, we plot both sides of Eq.~(\ref{equiltest}), which
are equal in equilibrium, for different ratios of the number of OR
sweeps to MET sweeps. The data is for $L=16, T=0.265$.
It is seen that equilibration is considerably
speeded up by including OR sweeps. It seems that doing 10 OR per
MET (which was used in the results in the earlier sections) is somewhat
better than 1 OR or 40 OR. Reference \onlinecite{campos:06} argues that
of order $L$ OR sweeps should be done for each MET sweep ``to let the
microcanonical wave run
over the system''. Our data is consistent with this, though it seems
that time to equilibrate is not very sensitive to the precise ratio of
OR to MET sweeps.

We should emphasize that including OR sweeps not only reduces the number
of sweeps to equilibrate, as seen in Fig.~\ref{opt_equil}, but also
reduces the CPU time by an even bigger
factor, because each OR sweep runs several times faster on the computer
than an MET sweep.

\section{Conclusions}
\label{sec:concl}
We have studied the XY spin glass in three dimensions by Monte Carlo
simulations using larger sizes than before. We find that the lower
critical dimension is close to three. We also find that the behavior of
the spin glass and chiral glass correlations lengths is strikingly
similar, see Fig.~\ref{xi_L_both}, and, in our view, does not support
the spin chirality decoupling scenario, at least for sizes up to $L=
24$.

In earlier work, Maucourt and Grempel\cite{maucourt:98} have studied the
3d XY spin glass using the domain-wall renormalization group (DWRG), for
sizes up to $L = 8$. They argue that there is a positive stiffness for
the chiralities, and hence a finite temperature transition, while for
spin glass ordering the system is close to its lower critical dimension.
The conclusion for chiralities is different from ours but we note
that our sizes are much larger ($L\le24$) and that we only see marginal
behavior in the chiralities for $L > 12$. Furthermore, our approach
gives \textit{directly} the correlation lengths, whereas for the DWRG
ground state energies with different boundary conditions are computed
from which a stiffness is \textit{inferred}.
%However, it is not very easy to
%understand what are the best boundary conditions from which to determine
%the stiffness, see e.g.~arXiv:cond-mat/9806339.

Kawamura and Li\cite{kawamura:01} used Monte Carlo simulations with
sizes up to $L=16$ to compute the overlap function of the spins and
chiralities. In particular, they compute the ``Binder ratio'' which,
like the ratio of the correlation length to system size studied here, is
dimensionless. The spin glass Binder ratio is found to monotonically
decrease with increasing $L$ at each temperature.
However, we feel that use of the Binder
ratio can be tricky near the lower critical dimension especially when
the number of components of the order parameter is high. Since the spin
glass order parameter is quadratic in the spins and the spins have two
components, the order parameter has four-components here. The Binder ratio
looks at the change in \textit{shape}
of the distribution of the (square root of the)
order parameter squared summed over all components, when going below the
transition. Because of the central limit theorem, there would be \textit{no}
change in shape for an infinite number of components. If the number is large
the change in shape is small and can easily be masked by corrections to
scaling, especially if the system is close to the lower critical dimension where
corrections
only fall off very slowly with system size.  The use of the Binder ratio
for vector spin glasses has also been criticized by Shirakura and
Matsubara\cite{shirakura:02} (they considered explicitly the Heisenberg case).
For the chiral glass Binder ratio, Kawamura and Li estimate a transition
temperature from a dip in the data. However, even if the transition is of an
unconventional kind (as they claim in order to explain the dip) it seems to us
that the Binder ratio should still increase with increasing $L$ at
low temperature if there is chiral glass order. However, this is not observed.

We therefore argue that our results, which compute \textit{directly} the
relevant correlation lengths, indicate that spin-chirality decoupling
does not seem to occur, at least for sizes up to $L = 24$.

Finally, we find that
equilibration is considerably speeded up by performing
several (perhaps of order $L$) over-relaxation sweeps per Metropolis
sweep, see Fig.~\ref{opt_equil}.

\begin{acknowledgments}
We acknowledge support
from the National Science Foundation under grant DMR 0337049 and are
also very grateful to the Hierarchical Systems Research Foundation for
a generous allocation of computer time on its Mac G5 cluster. We would
also would like to thank Helmut Katzgraber for helpful suggestions.
\end{acknowledgments}

\bibliography{refs}

\begin{thebibliography}{21}
\expandafter\ifx\csname natexlab\endcsname\relax\def\natexlab#1{#1}\fi
\expandafter\ifx\csname bibnamefont\endcsname\relax
  \def\bibnamefont#1{#1}\fi
\expandafter\ifx\csname bibfnamefont\endcsname\relax
  \def\bibfnamefont#1{#1}\fi
\expandafter\ifx\csname citenamefont\endcsname\relax
  \def\citenamefont#1{#1}\fi
\expandafter\ifx\csname url\endcsname\relax
  \def\url#1{\texttt{#1}}\fi
\expandafter\ifx\csname urlprefix\endcsname\relax\def\urlprefix{URL }\fi
\providecommand{\bibinfo}[2]{#2}
\providecommand{\eprint}[2][]{\url{#2}}

\bibitem[{\citenamefont{Ballesteros et~al.}(2000)\citenamefont{Ballesteros,
  Cruz, Fernandez, Martin-Mayor, Pech, Ruiz-Lorenzo, Tarancon, Tellez, Ullod,
  and Ungil}}]{ballesteros:00}
\bibinfo{author}{\bibfnamefont{H.~G.} \bibnamefont{Ballesteros}},
  \bibinfo{author}{\bibfnamefont{A.}~\bibnamefont{Cruz}},
  \bibinfo{author}{\bibfnamefont{L.~A.} \bibnamefont{Fernandez}},
  \bibinfo{author}{\bibfnamefont{V.}~\bibnamefont{Martin-Mayor}},
  \bibinfo{author}{\bibfnamefont{J.}~\bibnamefont{Pech}},
  \bibinfo{author}{\bibfnamefont{J.~J.} \bibnamefont{Ruiz-Lorenzo}},
  \bibinfo{author}{\bibfnamefont{A.}~\bibnamefont{Tarancon}},
  \bibinfo{author}{\bibfnamefont{P.}~\bibnamefont{Tellez}},
  \bibinfo{author}{\bibfnamefont{C.~L.} \bibnamefont{Ullod}}, \bibnamefont{and}
  \bibinfo{author}{\bibfnamefont{C.}~\bibnamefont{Ungil}},
  \emph{\bibinfo{title}{Critical behavior of the three-dimensional {I}sing spin
  glass}}, \bibinfo{journal}{Phys. Rev. B} \textbf{\bibinfo{volume}{62}},
  \bibinfo{pages}{14237} (\bibinfo{year}{2000}),
  \eprint{(arXiv:cond-mat/0006211)}.

\bibitem[{\citenamefont{Morris et~al.}(1986)\citenamefont{Morris, Colborne,
  Bray, Moore, and Canisius}}]{morris:86}
\bibinfo{author}{\bibfnamefont{B.~M.} \bibnamefont{Morris}},
  \bibinfo{author}{\bibfnamefont{S.~G.} \bibnamefont{Colborne}},
  \bibinfo{author}{\bibfnamefont{A.~J.} \bibnamefont{Bray}},
  \bibinfo{author}{\bibfnamefont{M.~A.} \bibnamefont{Moore}}, \bibnamefont{and}
  \bibinfo{author}{\bibfnamefont{J.}~\bibnamefont{Canisius}},
  \emph{\bibinfo{title}{Zero-temperature critical behaviour of vector spin
  glasses}}, \bibinfo{journal}{J. Phys. C} \textbf{\bibinfo{volume}{19}},
  \bibinfo{pages}{1157} (\bibinfo{year}{1986}).

\bibitem[{\citenamefont{Jain and Young}(1986)}]{jain:86}
\bibinfo{author}{\bibfnamefont{S.}~\bibnamefont{Jain}} \bibnamefont{and}
  \bibinfo{author}{\bibfnamefont{A.~P.} \bibnamefont{Young}},
  \emph{\bibinfo{title}{Monte {C}arlo simulations of {XY} spin-glasses}},
  \bibinfo{journal}{J. Phys. C} \textbf{\bibinfo{volume}{19}},
  \bibinfo{pages}{3913} (\bibinfo{year}{1986}).

\bibitem[{\citenamefont{Villain}(1977)}]{villain:77b}
\bibinfo{author}{\bibfnamefont{J.}~\bibnamefont{Villain}},
  \emph{\bibinfo{title}{Two-level systems in a spin-glass model. {I}. {G}eneral
  formalism and two-dimensional model}}, \bibinfo{journal}{J. Phys. C}
  \textbf{\bibinfo{volume}{10}}, \bibinfo{pages}{4793} (\bibinfo{year}{1977}).

\bibitem[{\citenamefont{Kawamura and Tanemura}(1987)}]{kawamura:87}
\bibinfo{author}{\bibfnamefont{H.}~\bibnamefont{Kawamura}} \bibnamefont{and}
  \bibinfo{author}{\bibfnamefont{M.}~\bibnamefont{Tanemura}},
  \emph{\bibinfo{title}{Chiral order in a two-dimensional {XY} spin glass}},
  \bibinfo{journal}{Phys. Rev. B} \textbf{\bibinfo{volume}{36}},
  \bibinfo{pages}{7177} (\bibinfo{year}{1987}).

\bibitem[{\citenamefont{Kawamura and Li}(2001)}]{kawamura:01}
\bibinfo{author}{\bibfnamefont{H.}~\bibnamefont{Kawamura}} \bibnamefont{and}
  \bibinfo{author}{\bibfnamefont{M.~S.} \bibnamefont{Li}},
  \emph{\bibinfo{title}{Nature of the ordering of the three-dimensional {XY}
  spin glass}}, \bibinfo{journal}{Phys. Rev. Lett.}
  \textbf{\bibinfo{volume}{87}}, \bibinfo{pages}{187204}
  (\bibinfo{year}{2001}), \eprint{(arXiv:cond-mat/0106551)}.

\bibitem[{\citenamefont{Maucourt and Grempel}(1998)}]{maucourt:98}
\bibinfo{author}{\bibfnamefont{J.}~\bibnamefont{Maucourt}} \bibnamefont{and}
  \bibinfo{author}{\bibfnamefont{D.~R.} \bibnamefont{Grempel}},
  \emph{\bibinfo{title}{Lower critical dimension of the {XY} spin-glass
  model}}, \bibinfo{journal}{Phys. Rev. Lett.} \textbf{\bibinfo{volume}{80}},
  \bibinfo{pages}{770} (\bibinfo{year}{1998}).

\bibitem[{\citenamefont{Akino and Kosterlitz}(2002)}]{akino:02}
\bibinfo{author}{\bibfnamefont{N.}~\bibnamefont{Akino}} \bibnamefont{and}
  \bibinfo{author}{\bibfnamefont{J.~M.} \bibnamefont{Kosterlitz}},
  \emph{\bibinfo{title}{Domain wall renormalization group study of {XY} model
  with quenched random phase shifts}}, \bibinfo{journal}{Phys. Rev. B}
  \textbf{\bibinfo{volume}{66}}, \bibinfo{pages}{054536}
  (\bibinfo{year}{2002}), \eprint{(arXiv:cond-mat/0203299)}.

\bibitem[{\citenamefont{Granato}(2000)}]{granato:00}
\bibinfo{author}{\bibfnamefont{E.}~\bibnamefont{Granato}},
  \emph{\bibinfo{title}{Phase-coherence transition in granular superconductors
  with $\pi$ junctions}}, \bibinfo{journal}{J. Magn. Magn. Matter.}
  \textbf{\bibinfo{volume}{226}}, \bibinfo{pages}{366} (\bibinfo{year}{2000}),
  \eprint{(arXiv:cond-mat/0012238)}.

\bibitem[{\citenamefont{Lee and Young}(2003)}]{leeLW:03}
\bibinfo{author}{\bibfnamefont{L.~W.} \bibnamefont{Lee}} \bibnamefont{and}
  \bibinfo{author}{\bibfnamefont{A.~P.} \bibnamefont{Young}},
  \emph{\bibinfo{title}{Single spin- and chiral-glass transition in vector spin
  glasses in three-dimensions}}, \bibinfo{journal}{Phys. Rev. Lett.}
  \textbf{\bibinfo{volume}{90}}, \bibinfo{pages}{227203}
  (\bibinfo{year}{2003}), \bibinfo{note}{(referred to as LY)},
  \eprint{(arXiv:cond-mat/0302371)}.

\bibitem[{\citenamefont{Katzgraber et~al.}(2006)\citenamefont{Katzgraber,
  K\"orner, and Young}}]{katzgraber:06}
\bibinfo{author}{\bibfnamefont{H.~G.} \bibnamefont{Katzgraber}},
  \bibinfo{author}{\bibfnamefont{M.}~\bibnamefont{K\"orner}}, \bibnamefont{and}
  \bibinfo{author}{\bibfnamefont{A.~P.} \bibnamefont{Young}},
  \emph{\bibinfo{title}{Detailed study of universality in three-dimensional
  {I}sing spin glasses}}, \bibinfo{journal}{Phys. Rev. B}
  \textbf{\bibinfo{volume}{73}}, \bibinfo{pages}{224432}
  (\bibinfo{year}{2006}), \eprint{(arXiv:cond-mat/0602212)}.

\bibitem[{\citenamefont{Hasenbusch et~al.}(2008)\citenamefont{Hasenbusch,
  Pellissetto, and Vicari}}]{hasenbusch:08}
\bibinfo{author}{\bibfnamefont{M.}~\bibnamefont{Hasenbusch}},
  \bibinfo{author}{\bibfnamefont{A.}~\bibnamefont{Pellissetto}},
  \bibnamefont{and} \bibinfo{author}{\bibfnamefont{E.}~\bibnamefont{Vicari}},
  \emph{\bibinfo{title}{The critical behavior of 3d {I}sing glass models:
  universality and scaling corrections}}, \bibinfo{journal}{J. Stat. Mech.} p.
  \bibinfo{pages}{L02001} (\bibinfo{year}{2008}),
  \bibinfo{note}{(arXiv:0710.1980)}.

\bibitem[{\citenamefont{Campos et~al.}(2006)\citenamefont{Campos, Cotallo-Aban,
  Martin-Mayor, Perez-Gaviro, and Tarancon}}]{campos:06}
\bibinfo{author}{\bibfnamefont{I.}~\bibnamefont{Campos}},
  \bibinfo{author}{\bibfnamefont{M.}~\bibnamefont{Cotallo-Aban}},
  \bibinfo{author}{\bibfnamefont{V.}~\bibnamefont{Martin-Mayor}},
  \bibinfo{author}{\bibfnamefont{S.}~\bibnamefont{Perez-Gaviro}},
  \bibnamefont{and} \bibinfo{author}{\bibfnamefont{A.}~\bibnamefont{Tarancon}},
  \emph{\bibinfo{title}{Spin-glass transition of the three-dimensional
  {H}eisenberg spin glass}}, \bibinfo{journal}{Phys. Rev. Lett.}
  \textbf{\bibinfo{volume}{97}}, \bibinfo{pages}{217204}
  (\bibinfo{year}{2006}).

\bibitem[{\citenamefont{Lee and Young}(2007)}]{lee:07}
\bibinfo{author}{\bibfnamefont{L.~W.} \bibnamefont{Lee}} \bibnamefont{and}
  \bibinfo{author}{\bibfnamefont{A.~P.} \bibnamefont{Young}},
  \emph{\bibinfo{title}{Large-scale {M}onte {C}arlo simulations of the
  isotropic three-dimensional {H}eisenberg spin glass}},
  \bibinfo{journal}{Phys. Rev. B} \textbf{\bibinfo{volume}{76}},
  \bibinfo{pages}{024405} (\bibinfo{year}{2007}),
  \eprint{(arXiv:cond-mat/0703770)}.

\bibitem[{\citenamefont{Hukushima and Kawamura}(2005)}]{hukushima:05}
\bibinfo{author}{\bibfnamefont{K.}~\bibnamefont{Hukushima}} \bibnamefont{and}
  \bibinfo{author}{\bibfnamefont{H.}~\bibnamefont{Kawamura}},
  \emph{\bibinfo{title}{Monte {C}arlo simulations of the phase transition of
  the three-dimensional isotropic {H}eisenberg spin glass}},
  \bibinfo{journal}{Phys. Rev. B} \textbf{\bibinfo{volume}{72}},
  \bibinfo{pages}{144416} (\bibinfo{year}{2005}).

\bibitem[{\citenamefont{Alonso et~al.}(1996)\citenamefont{Alonso,
  A.~Taranc\'on, Ballesteros, Fern\'andez, Mart\'in-Mayor, and Mu\~noz
  Sudupe}}]{alonso:96}
\bibinfo{author}{\bibfnamefont{J.}~\bibnamefont{Alonso}},
  \bibinfo{author}{\bibfnamefont{A.}~\bibnamefont{A.~Taranc\'on}},
  \bibinfo{author}{\bibfnamefont{H.}~\bibnamefont{Ballesteros}},
  \bibinfo{author}{\bibfnamefont{L.}~\bibnamefont{Fern\'andez}},
  \bibinfo{author}{\bibfnamefont{V.}~\bibnamefont{Mart\'in-Mayor}},
  \bibnamefont{and} \bibinfo{author}{\bibfnamefont{A.}~\bibnamefont{Mu\~noz
  Sudupe}}, \emph{\bibinfo{title}{Monte {C}arlo study of {O}(3)
  antiferromagnetic models in three dimensions}}, \bibinfo{journal}{Phys. Rev.
  B} \textbf{\bibinfo{volume}{53}}, \bibinfo{pages}{2537}
  (\bibinfo{year}{1996}).

\bibitem[{\citenamefont{Palassini and Caracciolo}(1999)}]{palassini:99b}
\bibinfo{author}{\bibfnamefont{M.}~\bibnamefont{Palassini}} \bibnamefont{and}
  \bibinfo{author}{\bibfnamefont{S.}~\bibnamefont{Caracciolo}},
  \emph{\bibinfo{title}{Universal finite size scaling functions in the 3d
  {I}sing spin glass}}, \bibinfo{journal}{Phys. Rev. Lett.}
  \textbf{\bibinfo{volume}{82}}, \bibinfo{pages}{5128} (\bibinfo{year}{1999}),
  \eprint{(arXiv:cond-mat/9904246)}.

\bibitem[{\citenamefont{Hukushima and Nemoto}(1996)}]{hukushima:96}
\bibinfo{author}{\bibfnamefont{K.}~\bibnamefont{Hukushima}} \bibnamefont{and}
  \bibinfo{author}{\bibfnamefont{K.}~\bibnamefont{Nemoto}},
  \emph{\bibinfo{title}{Exchange {M}onte {C}arlo method and application to spin
  glass simulations}}, \bibinfo{journal}{J. Phys. Soc. Japan}
  \textbf{\bibinfo{volume}{65}}, \bibinfo{pages}{1604} (\bibinfo{year}{1996}).

\bibitem[{\citenamefont{Marinari}(1998)}]{marinari:98b}
\bibinfo{author}{\bibfnamefont{E.}~\bibnamefont{Marinari}},
  \emph{\bibinfo{title}{Optimized {M}onte {C}arlo methods}}, in
  \emph{\bibinfo{booktitle}{Advances in Computer Simulation}}, edited by
  \bibinfo{editor}{\bibfnamefont{J.}~\bibnamefont{Kert\'esz}} \bibnamefont{and}
  \bibinfo{editor}{\bibfnamefont{I.}~\bibnamefont{Kondor}}
  (\bibinfo{publisher}{Springer-Verlag}, \bibinfo{year}{1998}),
  p.~\bibinfo{pages}{50}, \eprint{(arXiv:cond-mat/9612010)}.

\bibitem[{\citenamefont{Katzgraber et~al.}(2001)\citenamefont{Katzgraber,
  Palassini, and Young}}]{katzgraber:01}
\bibinfo{author}{\bibfnamefont{H.~G.} \bibnamefont{Katzgraber}},
  \bibinfo{author}{\bibfnamefont{M.}~\bibnamefont{Palassini}},
  \bibnamefont{and} \bibinfo{author}{\bibfnamefont{A.~P.} \bibnamefont{Young}},
  \emph{\bibinfo{title}{{M}onte {C}arlo simulations of spin glasses at low
  temperatures}}, \bibinfo{journal}{Phys. Rev. B}
  \textbf{\bibinfo{volume}{63}}, \bibinfo{pages}{184422}
  (\bibinfo{year}{2001}), \eprint{(arXiv:cond-mat/0007113)}.

\bibitem[{\citenamefont{Shirakura and Matsubara}(2003)}]{shirakura:02}
\bibinfo{author}{\bibfnamefont{T.}~\bibnamefont{Shirakura}} \bibnamefont{and}
  \bibinfo{author}{\bibfnamefont{F.}~\bibnamefont{Matsubara}},
  \emph{\bibinfo{title}{Binder parameter of a {H}eisenberg spin-glass model in
  four dimensions}}, \bibinfo{journal}{Phys. Rev. B}
  \textbf{\bibinfo{volume}{67}}, \bibinfo{pages}{100405}
  (\bibinfo{year}{2003}), \eprint{(arXiv:cond-mat/0211521)}.

\end{thebibliography}

\end{document}